\documentclass[3p,times,twocolumn]{elsarticle}

\usepackage{amssymb}

\usepackage[figuresright]{rotating}

\usepackage[pdftex,colorlinks=true,citecolor=blue,urlcolor=blue]{hyperref}

\begin{document}

\begin{frontmatter}




\title{Search for time modulations in the decay constant of $^{40}$K and $^{226}$Ra at the underground
Gran Sasso Laboratory}


\author[mi]{E. Bellotti}
\author[pd]{C. Broggini\corref{cor1}}
\author[lngs]{G. Di Carlo}
\author[lngs]{M. Laubenstein}
\author[pd]{R. Menegazzo}

\address[mi]{Universit\`{a} degli Studi di Milano Bicocca and Istituto Nazionale di Fisica Nucleare, Sezione di Milano, Milano, Italy}
\address[pd]{Istituto Nazionale di Fisica Nucleare, Sezione di Padova, Padova, Italy}
\address[lngs]{Istituto Nazionale di Fisica Nucleare, Laboratori Nazionali del Gran Sasso, Assergi (AQ), Italy}

\cortext[cor1]{Corresponding author}

\begin{abstract}
Time modulations at per mil level have been reported
to take place in the decay constant of several nuclei with period of one year (most cases) but also of about one month or one day.
On the other hand, experiments with similar or better sensitivity have been unable to detect any modulation. In this letter we give the results of the
activity study of two different sources:  $^{40}$K and $^{226}$Ra. The two gamma spectrometry experiments have been performed underground at the Gran Sasso Laboratory, this way
suppressing the time dependent cosmic ray background.
Briefly, our measurements reached the sensitivity of 3.4 and 3.5 parts over 10$^{6}$ for $^{40}$K and $^{226}$Ra, respectively (1 sigma) and they
do not show any statistically significant evidence of time dependence in the decay constant.
We also give the results of the activity measurement at the time of the two strong X-class solar flares which took place in September 2017. Our data do not show any unexpected time dependence in the decay rate of $^{40}$K in correspondence with the two flares.
To the best of our knowledge, these are the most precise and accurate results on the stability of the decay constant as function of time.
\end{abstract}

\end{frontmatter}


\section{Introduction}
A possible time dependence of the decay constant has already been discussed at the beginning of the science of radioactivity.
As a matter of fact, at the end of the Ph.D. thesis of Marie Curie one finds the description of the search, with negative results, for a difference in the activity of
uranium ores between midday and midnight.
Recently, in particular since 2009 \cite{Jenkins200942}, various experiments have reported evidence of a time modulation of the  decay constant of several  radioactive nuclei,
from $^{3}$H to $^{239}$Pu,
with period, in most cases, of one year (but also of about one month or one day) and amplitude at the per mil level \cite{Parkhomov2011,Jenkins201350,Sturrock201447}.
The annual modulation, with the maximum in February and the minimum in August, has been correlated to the change of the Sun-Earth distance between aphelion and perihelion.
In \cite{Fischbach2009,Jenkins2009407}  the interaction with solar neutrinos or the coupling to a long range scalar field from the Sun have been advocated as possible reason for the modulation. However, neutrino cross sections orders of magnitude higher than expected would be required.
In addition, the laboratory constraints on the variation of $\alpha_{em}$ and of the electron to proton mass ratio induce upper bounds to the relative variation of the decay constant nine orders of magnitude lower than the claimed per mil effect \cite{Bellotti201582}.

On the other hand, various experiments with similar or better sensitivity did not detect any modulation of the decay constants of several nuclei
(a complete review is in \cite{Pommé2016281}). Evidence against solar influences on nuclear decay constants in $\alpha$, $\beta^{-}$, $\beta^{+}$
and electron capture decays \cite{586793318,0026-1394-54-1-19,0026-1394-54-1-36} has been recently published using the data of 14 radionuclide metrology laboratories.

In the past we also performed a few gamma spectroscopy experiments in the underground Gran Sasso Laboratory, excluding modulations with
amplitude larger than a few parts over 10$^{5}$ in $^{137}$Cs \cite{Bellotti2012114}, $^{222}$Rn \cite{Bellotti2015526} and $^{232}$Th \cite{Bellotti201582}.

\section{The $^{40}$K experiment}
We already studied the decay constant of the electron capture decay of $^{40}$K with an over-ground experiment. To our knowledge, this remains the only study on
the subject. In particular, we measured a clear annual modulation with amplitude of 4.5$\pm$0.8$\cdot$10$^{-5}$ and maximum at
August 11th ($\pm$ 13 days). However, such an effect corresponds to a $\pm$ 3.5$\%$ modulation of the cosmic ray background (which we have measured) and it is well compatible,
both in size and in phase, with the known annual modulation of the cosmic
ray flux at the Earth surface (due to the change of the density in the upper atmosphere because of the temperature variation). In order to study the annual
modulation of the $^{40}$K decay independently of the comic ray flux, we built a new set-up which we installed deep underground in the Gran Sasso Laboratory, in a dedicated
container of 2m x 3.5m and 2.5m high placed in front
of Hall B. As a matter of fact, the mountain
shield suppresses the muon and neutron flux by six and three orders of magnitude, respectively,
as compared to the above ground.

\begin{figure}[th]
\centerline{
\includegraphics[width=0.5\textwidth,angle=0]{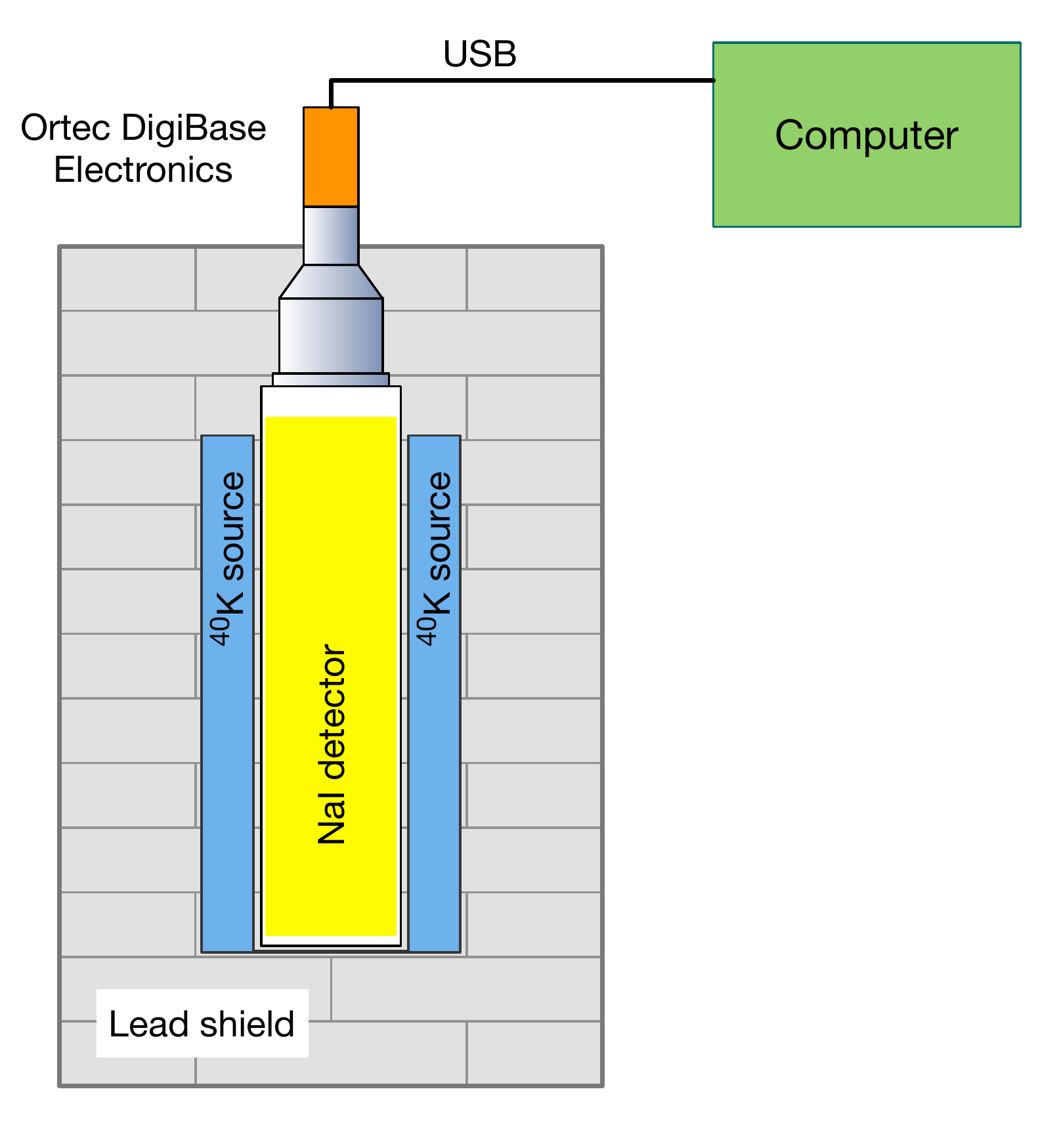}
}
\vskip -5mm
\caption{The $^{40}$K experimental set-up: the NaI crystal enclosed by the stainless steel box containing KHCO$_{3}$ melted with paraffine and shielded with 10 cm of lead.
}
\label{fig:setup40K}
\end{figure}

A 4 liter 4"x4"x16" NaI crystal detects the 1461 keV $\gamma$ ray due to the electron capture decay of $^{40}$K to the excited state of $^{40}$Ar (10.55 $\%$ branching ratio, half-life
1.25$\cdot$10$^{9}$ y).
The source is made of about 9.6 kg of potassium bicarbonate powder (KHCO$_{3}$, corresponding to 3.8 kg of natural potassium) and 2.4 kg of paraffine grains contained inside an
 hermetically sealed stainless steel box placed around the NaI detector.
The bicarbonate and the paraffine grains have been well mixed together before being poured inside the stainless steel box. The box has then been kept inside an oven
for about 20 hours at the temperature of 60 degrees Celsius. This way salt and paraffine melt together and, after cooling, a solid and compact block was formed. As a consequence
it was possible to avoid the slow and
long process of settlement of the salt powder (about 6 months) which was measured in  \cite{Bellotti201582}, where  KHCO$_{3}$
was used alone, and which was causing a time dependent detection efficiency.
Finally, the whole set-up is shielded by at least 10 cm of lead (Fig.~\ref{fig:setup40K}).

\begin{figure}[h]
\hskip 2mm
\centerline{
\includegraphics[width=0.52\textwidth,angle=0]{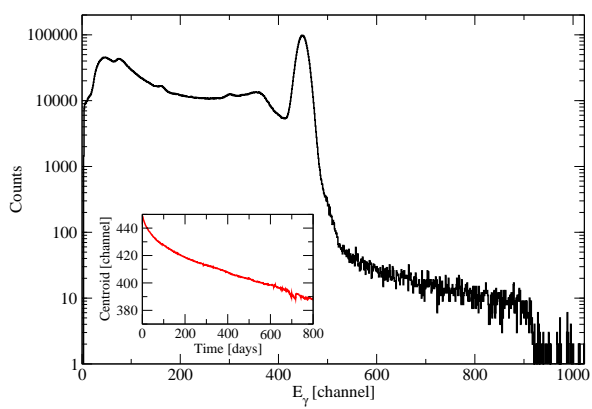}
}
\vskip -5mm
\caption{Measured $\gamma$-ray spectrum of the $^{40}$K source and centroid position of the 1461 keV peak during the measurement (in the inset).
}
\label{fig:histo40K}
\end{figure}

The electronic signals are processed by an Ortec (R) digiBASE (TM) with shaping time of 0.75 $\mu$s.
In the analysis we consider
the entire energy spectrum and not only the full energy peak at 1461 keV (Fig.~\ref{fig:histo40K}). This because we want to avoid any inaccuracy coming from the fitting procedure and we also want
to increase the total rate in order to improve the statistics. This procedure
requires the definition of lower and upper boundaries. Since
the content of the energy spectrum above $\sim$3 MeV is negligible, the only delicate
point is the stability of the lower threshold that should be low enough to
collect the entire spectrum and high enough to be well above the electronic noise.
In addition, if the low energy threshold is sufficiently low and placed inside a flat region of the spectrum then it is also less sensitive to the variation of the global gain of the electronic chain.
The intrinsic background, i.e shielded set-up without the KHCO$_{3}$ salt, has been measured during a period of 12 days. Thanks to the underground environment
and to the detector shielding, it is rather low, down to about 6.4 Hz above 15 keV.

Spectra are stored once per hour with a dead time of 1.24-1.25$\cdot$10$^{-2}$, which has a fluctuation of 9$\cdot$10$^{-6}$ (1 sigma).
The time when the system is unable to process input signals is essentially the ADC conversion time, i.e. the time required by the ADC to digitize the energy signals,
and the time required by the electronics to store the event in the internal memory. This dead time is internally calculated by the digiBASE system
and has been here normalized by the time that the spectrum was being recorded.
The timing for data acquisition is provided by the internal quartz
oscillator of the acquisition card. Its precision and stability
(better than 10 ppm/year) are enough for our purposes.
During the 799 days of data taking, from September 2015 to November 2017,
we observed a monotonic change of the energy conversion gain (inset of Fig.~\ref{fig:histo40K}): this produced a maximum shift of 61 channels for the $^{40}$K peak at 1461 keV energy
(initially at channel 449). The room temperature stayed within a 13.5-15 degree Celsius window.

\begin{figure}[h]
\centerline{
\includegraphics[width=0.6\textwidth,angle=0]{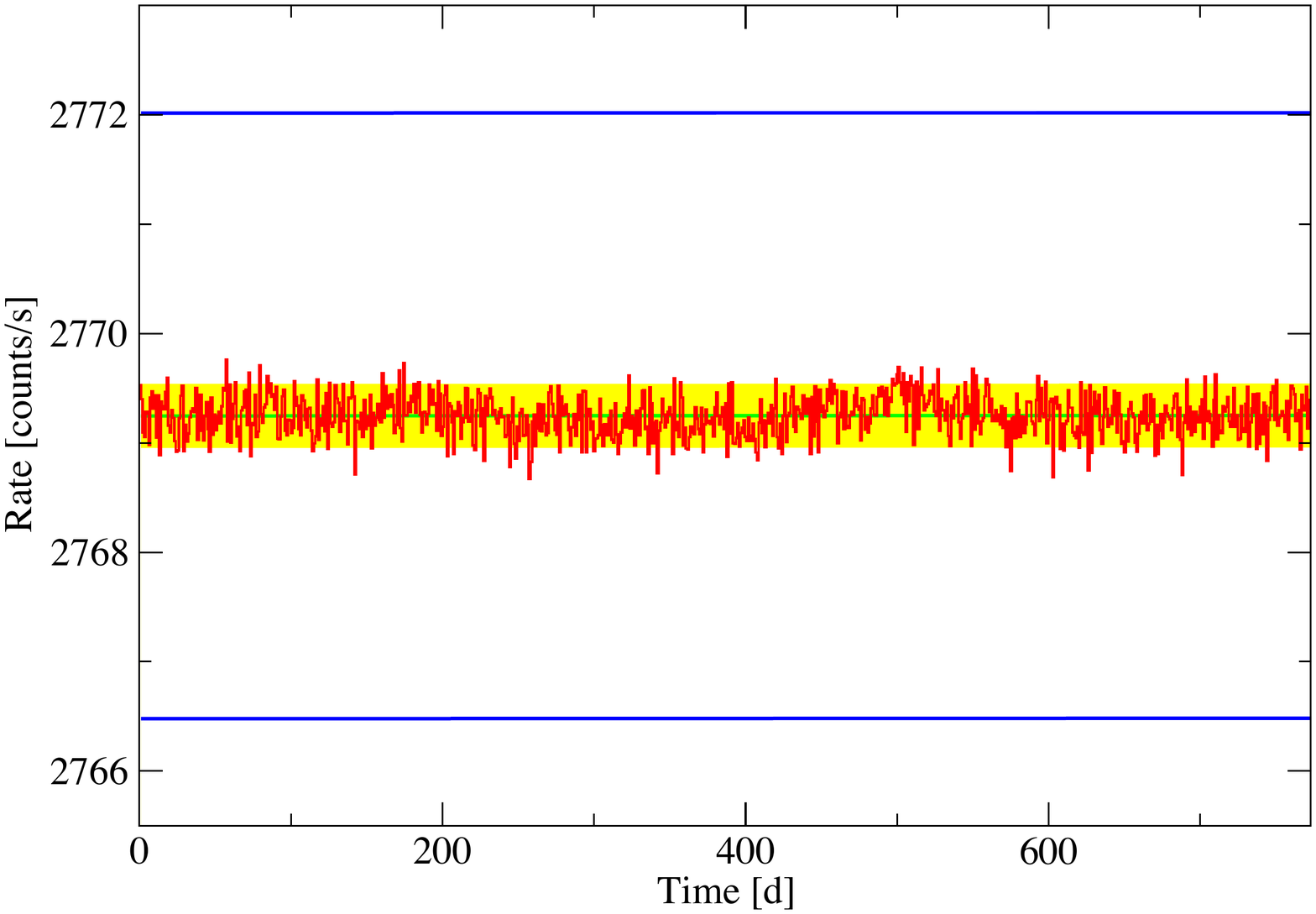}
}
\vskip -6mm
\caption{Measured rate of the $^{40}$K source averaged over 1 day. Horizontal
lines are drawn at $\pm$ 10$^{-3}$ from the average. The shaded area corresponds to a 10$^{-4}$ uncertainty.
}
\label{fig:rate40K}
\end{figure}

\begin{figure}[h]
\centerline{
\includegraphics[width=0.6\textwidth,angle=0]{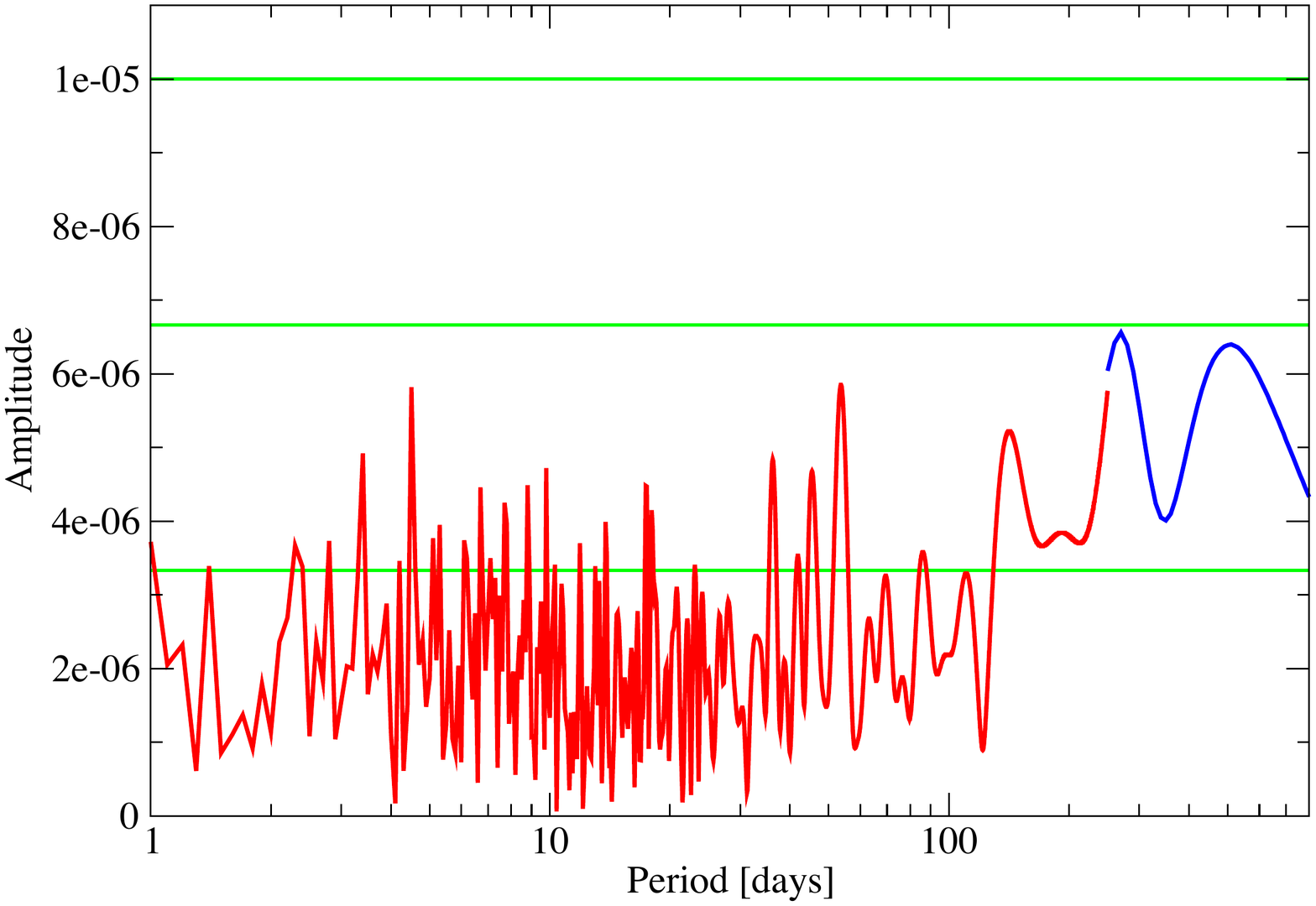}
}
\vskip -6mm
\caption{Potassium source: amplitude of the different periods from the Fourier transform of the residuals and from the chi-squared analysis. Horizontal lines correspond to 1, 2 and 3 standard deviations, obtained with a 10000 time re-shuffling of the residuals or from the fit parameters.
}
\label{fig:analysis40K}
\end{figure}

The rate of about 2770 Hz is shown in Fig.~\ref{fig:rate40K} as function of time.
We do not make any dead time correction to the data but we apply a correction to the
content of the first energy bin due to the peak position shift (at most 165000 counts/day, corresponding to 0.069 $\%$ of the rate).
The rate as function of time, when fitted with a constant and using only Poisson fluctuations as variance, gives a satisfactory chi squared per degree of freedom,
$\chi^2/dof$, of 1.11. At this point, the amplitudes of possible
time modulations of the rate are searched
for with the Fourier transform method (FFT) applied to the residuals or with the minimization of the chi squared fit of the residuals with a cosine function of time. As a matter of fact, Fourier transform can be rigorously applied only when searching for periods
significantly smaller than the counting time. In particular, we have chosen a limit equal to one third of the run time. Finally,
the statistical significance of the different amplitudes is obtained with a 10000 time re-shuffling of the residuals
for the FFT analysis or directly from the errors on the fit parameters
for the chi-squared analysis.
Fig.~\ref{fig:analysis40K} shows the results of the Fourier transform of the residuals, for periods from 1 to 250 days, and of the chi-squared
analysis for longer periods.

In particular, using the chi-squared analysis of the residuals fitted with a cosine function of time we obtain the amplitude
of 4.2$\pm$3.3$\cdot$10$^{-6}$ (1 sigma) for the period of 1 year and phase unconstrained. The maximum amplitude provided by the chi-squared analysis
is 6.6$\pm$3.3$\cdot$10$^{-6}$ (1 sigma) for a period of 270 days and phase unconstrained. Both the amplitudes are not statistically
significant.

\begin{figure}[h]
\centerline{
\includegraphics[width=0.6\textwidth,angle=0]{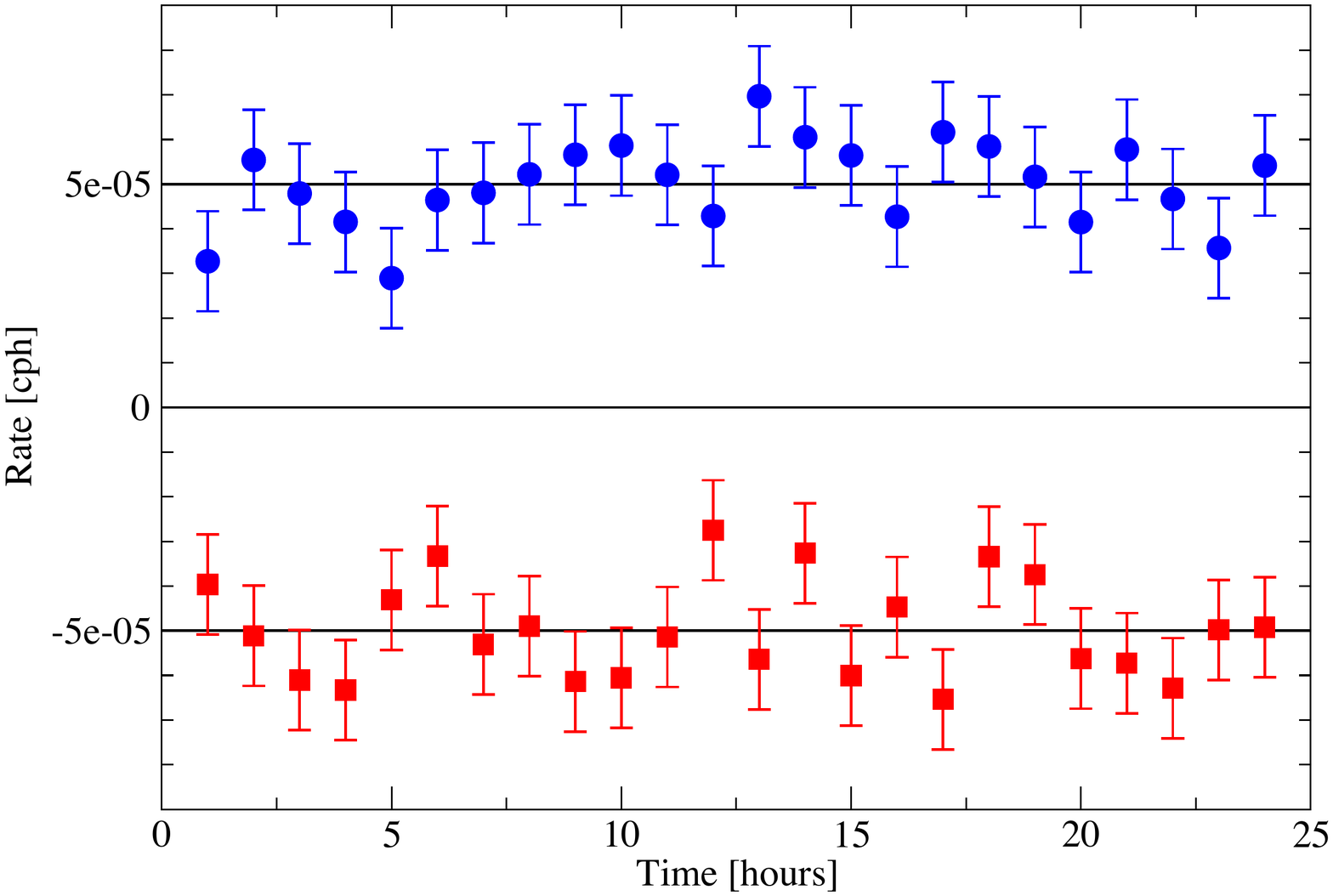}
}
\vskip -6mm
\caption{Normalized residuals (averaged over the hour) of the $^{40}$K source as function of solar (dots) and sidereal (squares) hours. Statistical uncertainties are indicated. An offset $\pm$5 $\cdot$ 10$^{-5}$ has been added to the datasets to improve the clarity.
}
\label{fig:daily40K}
\end{figure}

Finally, we have studied the rate as function of the 24 hours of the solar and sidereal time
(Fig.~\ref{fig:daily40K}) to search for a possible daily modulation (i.e. 24 hour period and unknown phase). Both the rates are well compatible with a constant
($\chi^2/dof$ of 0.76 and 1.00, respectively) and time modulation amplitudes larger than 2.0 $\cdot$ 10$^{-6}$ and 2.3 $\cdot$ 10$^{-6}$ are
excluded for the solar and sidereal time, respectively, by the chi-squared analysis of the rate with a constant fit.

\section{The $^{226}$Ra experiment}
The set-up is installed in the same laboratory as the $^{40}$K experiment.
The source is a $^{226}$Ra standard source
($\alpha$ decay + $\alpha$/$\beta$ decay of the chain)
with Radium kept inside a sealed glass tube embedded in a plastic disk of 1" diameter and 1/8" thick. All the Radon produced
from the decay of $^{226}$Ra stays inside the glass tube, thus preserving the saecular equilibrium. The source is kept at fixed distance from the flat face of
the detector, a
3"x3" NaI crystal horizontally placed.

\begin{figure}[h]
\centerline{
\includegraphics[width=0.6\textwidth,angle=0]{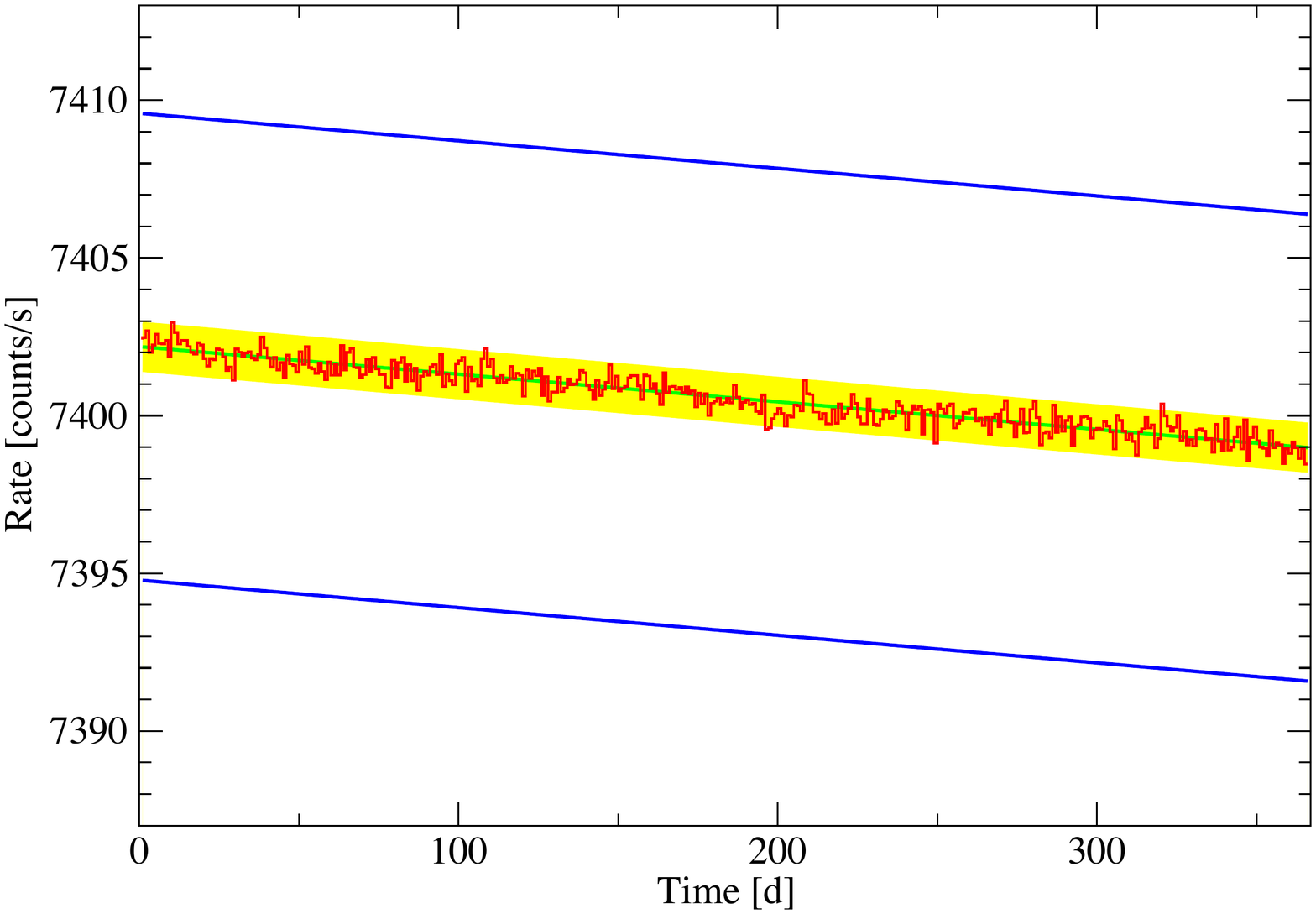}
}
\vskip -6mm
\caption{Measured rate of the $^{226}$Ra source averaged over 1 day. The two
lines are drawn at $\pm$ 10$^{-3}$ from the average. The shaded area corresponds to a 10$^{-4}$ uncertainty.
}
\label{fig:rate226Ra}
\end{figure}

Also in this experiment the signals are processed by an Ortec (R) digiBASE (TM) with shaping time of 0.75 $\mu$s
and the entire energy spectrum is used to obtain the rate of the $^{226}$Ra source, 7170 Hz above the 10 keV threshold.
The intrinsic background has been measured during a period of 12 days. It amounts to 1.01 Hz above 10 keV.

Gamma spectra are stored every 3 hours with a dead time of 3.18$\cdot$10 $^{-2}$, which has a fluctuation of
4$\cdot$10$^{-5}$.
During the 1 year running time, from November 2015 to November 2016, the rate decreased by 0.032 $\%$, as expected from the 1600 y half-life of  $^{226}$Ra,  the dead time decreased to
3.16$\cdot$10 $^{-2}$
and the peak at 1120 keV energy monotonically shifted by
39 channels. Because of this shift, we apply a correction to the
content of the first energy bin (0.08 $\%$ at most).
The rate as function of time, corrected for the dead time, is shown in Fig.~\ref{fig:rate226Ra}.
When fitted with the expected decay law and using only Poisson fluctuations as variance, it gives a
$\chi^2/dof$ of 1.32.

\begin{figure}[h]
\centerline{
\includegraphics[width=0.6\textwidth,angle=0]{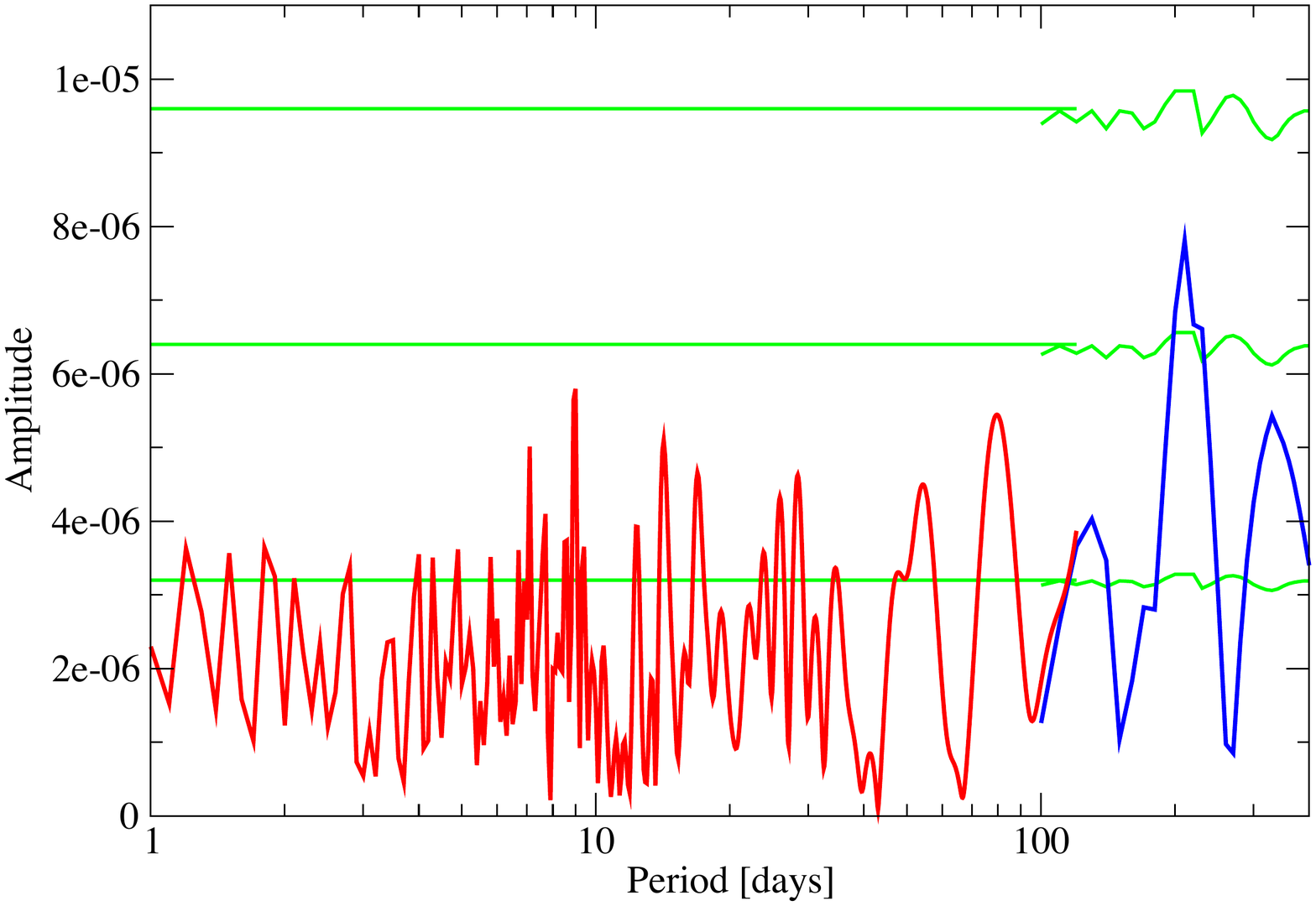}
}
\vskip -6mm
\caption{Radium source: amplitude of the different periods from the Fourier transform of the residuals and from the chi-squared
analysis. Horizontal lines correspond to 1, 2 and 3 standard deviations, obtained with a 10000 time re-shuffling of the residuals or from the fit parameters.
}
\label{fig:analysis226Ra}
\end{figure}

Fig.~\ref{fig:analysis226Ra} shows the results of the Fourier transform of the residuals, for periods from 1 to 100 days, and of the chi-squared
analysis for longer periods.

In particular, using the chi-squared analysis of the residuals fitted with a cosine function of time we obtain the amplitude
of 4.7$\pm$3.2$\cdot$10$^{-6}$ (1 sigma) for the period of 1 year and phase unconstrained. The maximum amplitude provided by the chi-squared analysis
is 7.8$\pm$3.3$\cdot$10$^{-6}$ (1 sigma) for a period of 210 days and phase unconstrained. Both the amplitudes are not statistically
significant.
We note that our sensitivity is a factor 300 smaller than the $^{226}$Ra modulation discussed in \cite{Jenkins200942} and one order of magnitude smaller than the best sensitivity
quoted in \cite{Pommé2016281}.

\section{Source activity during the two X-class solar flares of September 2017}
Solar flares are explosions on the Sun that happen when energy stored in twisted magnetic fields (usually above sunspots) is suddenly released.
This energy, up to one hundredth of the solar luminosity, is released within a few minutes to tens
of minutes.
Solar flares are classified according to the power of the X-ray flux peak near the Earth as
measured by the GOES-15 geostationary satellite:
X identifies the class of the most powerful ones, with a power at the peak larger than 10$^{-4}$  W/m$^{2}$
(within the X-class there is then a linear scale).

In \cite{Jenkins2009407} a significant dip (up to 4$\cdot$10$^{-3}$, $\sim$7 $\sigma$ effect) in the count rate,
averaged on a time interval of 4 hours, has been observed in the activity of a
$\sim$1 $\mu$Ci source of $^{54}$Mn (electron-capture) in coincidence with the X-3 and X-1 solar flares from December 2nd 2006 to January 1st 2007.

We already measured the decay constant during the occurrence of strong X-class solar
flares in the years 2011-2013. No significant deviations from expectations have been observed larger than a few parts over 10$^{4}$ \cite{Bellotti201582,Bellotti2013116}.
Here we give the results obtained by studying the activity of the $^{40}$K source in correspondence with the two X-class solar flares which took place on the 6th and the 10th of September 2017 (X9.3 and X8.2, respectively,
the most intense flares recorded during the current 11-year solar cycle, started on December 2008).

\begin{figure}[h]
\centerline{
\includegraphics[width=0.6\textwidth,angle=0]{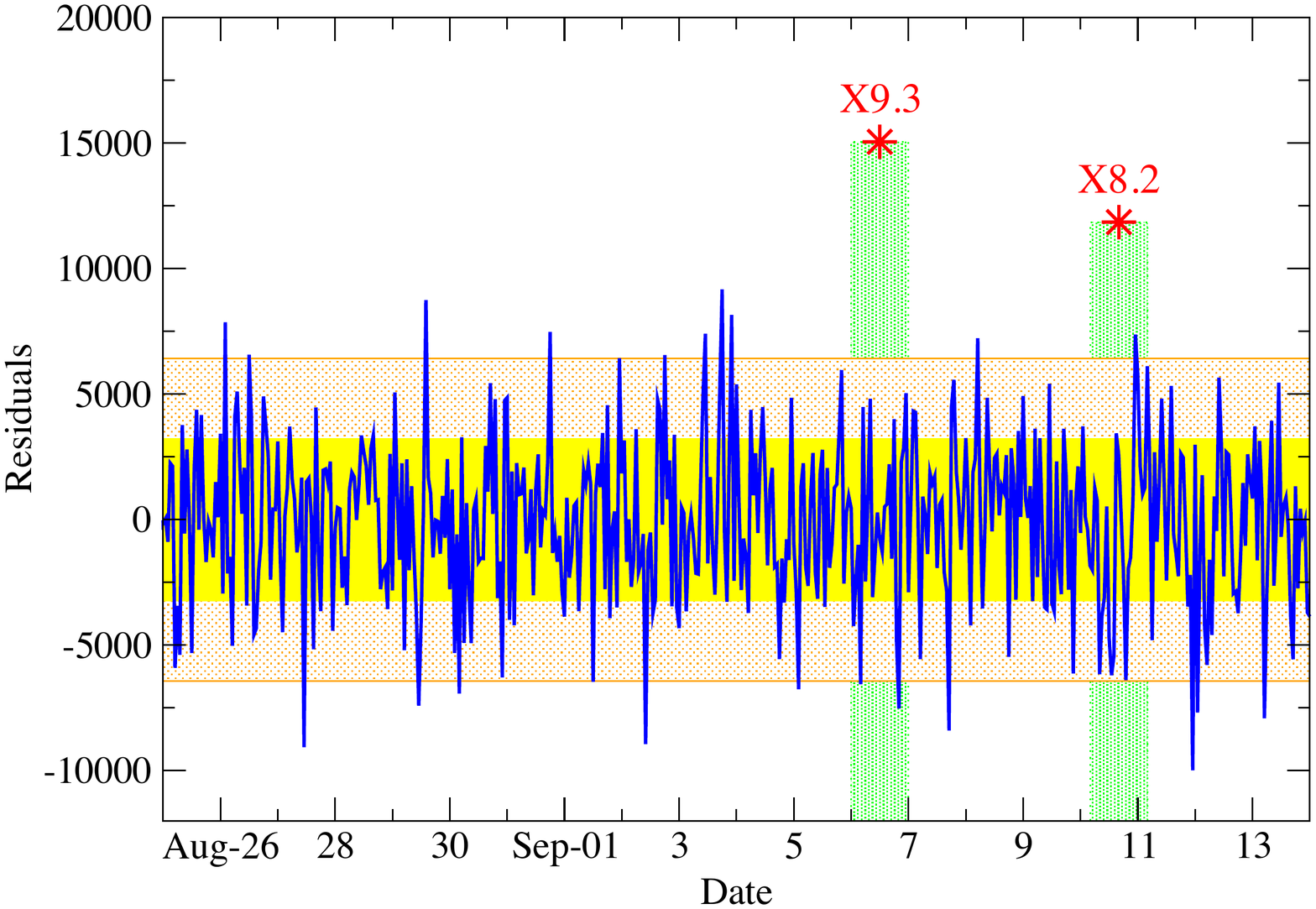}
}
\vskip -6mm
\caption{Residuals, averaged over 1 hour, of $^{40}$K data collected from the 25th of August 2017.
Shaded areas correspond to 1 and 2 standard deviations from the expected value.
}
\label{fig:flares}
\end{figure}

Fig.~\ref{fig:flares} shows the data collected in a 20 day window starting on the 25th of August 2017.
In particular, the residuals
of the count rate of the  $^{40}$K source are plotted, averaged over a period of 1 hour.
The error bands are purely statistical since systematic uncertainties are negligible as compared to the statistical ones.

Our data clearly exclude an effect as large as the one reported in \cite{Jenkins2009407}, of the order of a few per mil per day and lasting
several days. In particular,
the maximum effect compatible with our data is smaller than 6 $\cdot 10^{-5}$ per day and 3 $\cdot 10^{-4}$ per hour.

\section{Conclusion}
We have performed two gamma spectroscopy experiments in the underground Gran Sasso Laboratory to study the decay constant as function of time of $^{40}$K (electron capture decay) and
$^{226}$Ra ($\alpha$ decay + $\alpha$/$\beta$ decay of the chain). Thanks to the deep underground location it has been possible to remove the time dependent cosmic ray background. Our measurements have a sensitivity of 3.4 ($^{40}$K) and 3.5 ($^{226}$Ra) parts over 10$^{6}$ and they  do not show any evidence of time dependence
statistically significant in the decay constant.
We studied also the daily dependence of the $^{40}$K  decay constant on the solar and sidereal time: again no effect has been detected with a
limit on the relative amplitude of 2.0 $\cdot$ 10$^{-6}$  (solar time) and 2.4 $\cdot$ 10$^{-6}$ (sidereal time).
Finally, we exclude effects on the decay constant of $^{40}$K of relative amplitude larger than 6 $\cdot 10^{-5}$ per day and 3 $\cdot 10^{-4}$ per hour during the two solar flares of September 2017, the most powerful flares of the current 11-year solar cycle.

\section *{Bibliography}


\end{document}